\documentclass[10pt, conference, compsocconf]{IEEEtran}

\usepackage{cite}
\usepackage{url}
\usepackage{graphicx}
\usepackage{color}
\usepackage{xspace}

\newif\ifdraft
\drafttrue
\ifdraft
\newcommand{\comment}[1]{\textcolor{red}{#1}}
\newcommand{\vbnote}[1]{ {\textcolor{blue} {*VB: #1 }}}
\newcommand{\ehnote}[1]{ {\textcolor{green} {*EH: #1 }}}
\newcommand{\jhanote}[1]{ {\textcolor{cyan} {*SJ: #1 }}}
\else
 \newcommand{\comment}[1]{}
 \newcommand{\vbnote}[1]{}
 \newcommand{\ehnote}[1]{}
 \newcommand{\jhanote}[1]{}
\fi

\newcommand{\rp}{RADICAL-Pilot\xspace}
\newcommand{\enmt}{Ensemble Toolkit\xspace}
\newcommand{\ext}{ExTASY\xspace}

\begin{document}

\title{ExTASY: Scalable and Flexible Coupling of MD Simulations and Advanced
  Sampling Techniques}

\author{
\IEEEauthorblockN{
Vivekanandan Balasubramanian\IEEEauthorrefmark{1},
Iain Bethune\IEEEauthorrefmark{2},
Ardita Shkurti\IEEEauthorrefmark{3},
Elena Breitmoser\IEEEauthorrefmark{2}}
\IEEEauthorblockN{
Eugen Hruska\IEEEauthorrefmark{5}\IEEEauthorrefmark{6}, 
Cecilia Clementi\IEEEauthorrefmark{4}\IEEEauthorrefmark{6},
Charles Laughton\IEEEauthorrefmark{3} and
Shantenu Jha\IEEEauthorrefmark{1}}

\IEEEauthorblockA{ \IEEEauthorrefmark{1}Department of Electrical and Computer Engineering, 
Rutgers University, Piscataway, NJ 08854, USA}
\IEEEauthorblockA{\IEEEauthorrefmark{2}EPCC, The University of Edinburgh, James
Clerk Maxwell Building, Peter Guthrie Tait Road, Edinburgh, UK, EH9 3FD} 
\IEEEauthorblockA{\IEEEauthorrefmark{3} School of Pharmacy and Centre for
Biomolecular Sciences, University of Nottingham, University Park, Nottingham NG7 2RD, UK} 
\IEEEauthorblockA{\IEEEauthorrefmark{4} Department of Chemistry, Rice
University, Houston, TX 77005, USA} 
\IEEEauthorblockA{\IEEEauthorrefmark{5} Department of Physics, Rice
University, Houston, TX 77005, USA} 
\IEEEauthorblockA{\IEEEauthorrefmark{6} Center for Theoretical Biological
Physics, Rice University, Houston, TX 77005, USA} 
}

\maketitle
\thispagestyle{plain}
\pagestyle{plain} 

\begin{abstract}
  For many macromolecular systems the accurate sampling of the relevant regions
  on the potential energy surface cannot be obtained by a single, long Molecular
  Dynamics (MD) trajectory. New approaches are required to promote more
  efficient sampling.  We present the design and implementation of the
  Extensible Toolkit for Advanced Sampling and analYsis (ExTASY) for building
  and executing advanced sampling workflows on HPC systems.  ExTASY provides
  Python based ``templated scripts'' that interface to an interoperable and
  high-performance pilot-based run time system, which abstracts the complexity
  of managing multiple simulations.  ExTASY supports the use of existing
  highly-optimised parallel MD code and their coupling to analysis tools based
  upon collective coordinates which do not require a priori knowledge of the
  system to bias. We describe two workflows which both couple large
  ``ensembles'' of relatively short MD simulations with analysis tools to
  automatically analyse the generated trajectories and identify molecular
  conformational structures that will be used on-the-fly as new starting points
  for further ``simulation-analysis'' iterations.  One of the workflows
  leverages the Locally Scaled Diffusion Maps technique; the other makes use of
  Complementary Coordinates techniques to enhance sampling and generate
  start-points for the next generation of MD simulations.  We show that the
  ExTASY tools have been deployed on a range of HPC systems including ARCHER
  (Cray CX30), Blue Waters (Cray XE6/XK7), and Stampede (Linux cluster), and
  that good strong scaling can be obtained up to 1000s of MD simulations,
  independent of the size of each simulation.  We discuss how ExTASY can be
  easily extended or modified by end-users to build their own workflows, and
  ongoing work to improve the usability and robustness of ExTASY.
\end{abstract}

\section{Introduction and Motivation}

Approximately 30-40\% of compute cycles on US XSEDE~\cite{xsede30,xdmod_paper}
is devoted to research on biomolecular systems using Molecular Dynamics (MD)
simulations. Much of the computational cost comes from the need for an adequate
sampling of the conformational space accessible to these complex and flexible
systems in order to answer a particular research question.  For example, to
calculate free energies one needs an adequate sample from the Boltzmann weighted
ensemble of states for the system in order to estimate the thermodynamic
quantity of interest. Another example is the study of kinetic processes such as
self-assembly or drug-target association, where the integration of data from
large numbers of trajectories is required to build a statistically meaningful
model of the dynamical process.

The high dimensionality of these macromolecular systems and the complexity of
the associated potential energy surfaces (creating multiple metastable regions
connected by high free energy barriers) pose significant challenges to
adequately sample the relevant regions of the configurational space. In other
words, beside the ``curse of dimensionality'' associated with the large number
of degrees of freedom, MD trajectories can easily get ``trapped'' in a low free
energy state and fail to explore other biologically relevant states. The
waiting time to escape from a local free energy minimum increases exponentially
with the height of the free energy barrier that needs to be crossed to reach
another state. Metastable states separated by free energy barriers of several
tens of $k_B T$ (where $k_B$ is the Boltzmann constant and $T$ is the
physiological temperature) are not uncommon in biologically relevant systems,
but can not at present be routinely sampled with standard MD simulations.

In practice, better sampling of the relevant regions of a macromolecule
configurational space can be achieved through methodologies able to bias the
sampling towards scarcely visited regions, reducing the waiting time inside a
metastable state by artificially flattening the energy barriers between states
- e.g. Metadynamics \cite{Barducci2011} or Accelerated Dynamics
\cite{Pierce2012}. Although the results can be usually reweighed to reproduce
the correct Boltzmann statistics, kinetics properties are not easily recovered
from biased simulations (unless used in combinations with unbiased simulations,
see e.g.~\cite{Hu2016}). In addition, the design of an effective bias usually
requires some \textit{a priori} information on the system of interest, for
instance on a suitable choice of collective variables to describe slow
timescale processes. 

An alternative approach to tackle the sampling problem is the development of
ensemble or swarm simulation strategies, where data from large numbers of
simulations, which may be weakly coupled or not coupled at all, are integrated
(e.g. Replica Exchange \cite{Sugita1999} and Markov State Models (MSMs)\cite{Noe2014}).

This last class of methods is of increasing interest for a variety of reasons.
Firstly, the hardware roadmap is now based almost entirely on increasing core
counts, rather than clock speeds. On the face of it, these developments favour
weak scaling problems (larger and larger molecular systems to be simulated)
over strong scaling problems (getting more data faster on a system of fixed
size). However, by running ensembles of simulations over these cores and
integrating the data using, e.g. MSM approaches, timescales far
in excess of those sampled by any individual simulation are effectively accessed. 
In the last few years several studies have been published
\cite{Noe2009,Voelz2010,Buch2011,Huang2014,Plattner2015} where, using MSM
methods, processes such as protein folding or ligand binding have been
completely and quantitatively characterized (thermodynamically and kinetically)
from simulations orders of magnitude shorter than the process of interest. 

It is becoming increasingly clear that the application of ensemble simulation
strategies on state-of-the-art computational facilities has an unparalleled
potential to permit accurate simulations of the largest, most challenging, and
generally most biologically relevant, biomolecular systems. The main challenge 
in the development of the ensemble approach for faster sampling of complex
macromolecular systems is on the design of strategies to adaptively distribute
the trajectories over the relevant regions of the system’s configurational
space, without using any \textit{a priori} information on the system global
properties. The definition of smart ``adaptive sampling'' approaches that can
redirect computational resources towards unexplored yet relevant regions is
currently of great interest.

In light of the challenge posed by trends in computer architecture, the need to
improve sampling, and the range of existing MD codes and analysis tools, we have
designed and implemented the Extensible Toolkit for Advanced Sampling and
analYsis (ExTASY). ExTASY provides three key features within a single framework
to enable the development and applications requiring advanced sampling in a
flexible and highly scalable environment.  Firstly, as an extensible toolkit,
ExTASY allows a wide range of existing software to be integrated, leveraging the
significant community investment in highly optimised and well-tested software
packages and enabling users to continue to work with tools that they are
familiar with.  Support for specific MD codes and analysis tools is provided in
order to demonstrate how ExTASY may be used, but users can easily add other
tools as needed.

Secondly, ExTASY is flexible, providing a programmable interface to link
individual software components together and construct sampling workflows.
Workflows capture a sequence of execution of individual tools, and the data
transfers and dependencies between them.  First class support is provided for
defining large ensembles of independent simulations.  Thus complex calculations
may be scripted and then executed without the need for user intervention.

Thirdly, ExTASY workflows may be executed either locally, or on remote High
Performance Computing systems.  Complexities such as the batch queueing system
and data transfer are abstracted, making it easy for users to make use of the
most appropriate compute resources they have access to.  In addition, this
abstraction allows workflows to be executed without exposing each component to
queue waiting time, and respecting the dependencies defined in the workflow, for
many simulations to be scheduled and executed in parallel.

The rest of the paper is organized as follows: In the next section we discuss
the design and implementation of ExTASY. After a brief discussion in Section
III of two distinct applications that have been used to design and validate
ExTASY, Section IV provides a careful analysis of the performance and
scalability of ExTASY. Given the complex interplay between functionality and
performance when designing an extensible and production tool, we perform a wide
range of experiments aimed to investigate strong and weak scaling properties,
inter alia over a set of heterogeneous HPC platforms. We conclude with a
discussion of the scientific impact as well as the lessons for sustainable
software development.

\section{Related Work}

The need for better sampling has driven developments in methodology (algorithm),
hardware, and software for (bio)molecular simulation. 

One of the features of the popular Metadynamics method \cite{Barducci2011} or
Accelerated Dynamics \cite{Pierce2012} is that a constant analysis of what has
been sampled so far is used to bias future sampling into unexplored regions. A
range of alternative approaches are now emerging that do likewise, but where
alternating segments of data-gathering, and analysis to inform the direction of
future sampling, are more coarsely grained. This iterative approach has the
advantage over the metadynamics method that the identity of ``interesting''
directions for enhanced sampling does not need to be defined \textit{a priori},
but can emerge and respond flexibly to the developing ensemble. Another
advantage is that the MD-based sampling process and analysis method
do not have to be implemented within the same executable, or in two
tightly-coupled executables, permitting greater flexibility. Many such methods
make use of collective variables (CVs) to define directions in which to promote
sampling. A variety of novel, and established, algorithms for the unsupervised
and adaptive construction of CVs exist.  In addition to the work of Preto and
Clementi \cite{Preto2014}, interleaving cycles of MD simulation with data
analysis through Locally Scaled Diffusion Maps \cite{Rohrdanz2011}, related
methods include the non-targeted PaCS-MD method of Harada and Kitao
\cite{Guo2009}, variants thereof \cite{Harada2015}, and the PCA-based method of
Peng and Zhang \cite{Peng2014}.

Better sampling can also come from faster sampling, which has been enabled through
hardware developments such as ANTON \cite{Shaw:2008} and MD-GRAPE
\cite{Ohmura2013}. These special purpose computers enable much faster
calculations of the different contributions to the forces along the
trajectories, thus speeding up the clock time required to perform a time
integration step in the MD simulation and allowing execution of significantly
longer MD trajectories. The cost of, and access to such special purpose
computers ensure that in spite of their potential, they will not be as
accessible for the wider scientific community as general purpose approaches.
Further ANTON requires a customized ecosystem, from bespoke MD engines and ANTON
specific data analysis middleware (e.g., HiMach). Thus ANTON style
special-purpose approaches to bio-molecular simulation science cannot take
advantage of the rich community driven advances and eco-system.

Methods such as Replica Exchange and Metadynamics require a tight coupling
between the simulation and analysis processes, and are thus typically
implemented as additional facilities within the core MD code (e.g.  replica
exchange methods are implemented in AMBER~\cite{AMBER2013},
CHARMM~\cite{CHARMM1983}, GROMACS~\cite{GROMACS2015}, LAMMPS~\cite{LAMMPS1995},
and NAMD~\cite{NAMD2005}, for example), or are provided by a separate package
that communicates in a fine-grained manner with the running MD executable,
generally though specially-installed ``hooks''; an example of this approach is
the PLUMED package~\cite{PLUMED} which provides metadynamics capabilities
(amongst others) to AMBER, GROMACS, LAMMPS, NAMD and also Quantum
ESPRESSO~\cite{QE2009}.  In contrast, there is, to our knowledge, so far no
established and generally available package to support the types of
coarser-grained, adaptive workflows described above.

\section{ExTASY: Requirements, Design and Implementation}\label{sec:soft_arch}

In this section we first present the requirements that have been considered in
the design and implementation of ExTASY, which we then go on to discuss.

\subsection{Requirements}

Consistent with the design of many new software systems and tools, we analyze
the functionality, performance and usability requirements of ExTASY.

\subsubsection{Functionality}
Specific to sampling, there is a need to couple two very distinct computational
stages; each stage can be short-lived when compared to the typical duration of
monolithic simulation. Furthermore, the two stages differ in their resource
requirements significantly: one stage is characterized by multiple compute
intensive MD simulations, the other by a single analysis program that operates
on data aggregated from multiple simulations.  The ``Ex'' in ExTASY is a
reference to the extensible nature of the framework, and thus any coupling must
be between the abstraction of stages and not specific codes for fixed time
duration.

Scientists may have access to multiple systems, and wish to submit jobs on each,
or may be forced to migrate from using one system to another due to CPU time
allocation, or system end-of-life.  It is imperative that any software system
support \textbf{interoperability}, i.e. use of heterogeneous systems with
minimal changes. MD simulations may be executed over multiple nodes, depending
on the system size, thus any software system should also support the
\textbf{ability to execute tasks} over multiple nodes.

\subsubsection{Performance}
In order to obtain good sampling, a large number of simulations must be
executed. In many cases, the aggregate number of cores required by the set of
simulations (``ensemble'') is much higher than the total number of cores that
are available at a given instance or that can be allocated on a system. The
framework must decouple the aggregate (or peak) resource requirement of the
workload from the number of cores available or utilized.  On the other hand, if
access to a large number of cores is available, the framework should be able to
use them effectively. In this regard, the \textbf{strong and weak scalability}
of the framework is to be investigated.

\subsubsection{Usability}
Depending on the application, the user might need to change to using a
larger/smaller set of input data, modify simulation parameters or replace any of
the simulation or analysis tools. The framework should offer \textbf{easy
  application setup}, minimizing the user's time and effort in the process.  The
user should only be concerned with decisions on ``what" the workflow is and
``where" it is being executed. The details as to ``how" the deployment and
execution occurs should be hidden by the underlying software. Thus the framework
should use tools that \textbf{abstract complexities} of deployment and execution
from the user.  Workflow users and developers should be able to concentrate
their efforts on the application logic and expect the underlying software to
provide a level of transparent and automation of aspects such as data movement
and job submission.

\subsection{Design}

From these requirements, we identify the following as the primary design
objectives of \ext.

\begin{enumerate}
\item Support range of HPC systems, abstract task execution, data
  movement from the user.
\item Flexible resource management and coupling capabilities between different
  stages as well as within a stage.
\item Provide the users with an easy method to specify or change workload
  parameters without delving into the application itself.
\end{enumerate}

These design objectives put together lead to the following simple software architecture:

\subsubsection{Middleware for resource and execution management} The \ext
framework is aimed to provide an easy method to run advanced sampling algorithms
on HPC systems. In this process, the \ext framework should abstract users from
the complexity of application composition, workload execution and resource
management. There is need for a middleware that provides such resource and
execution management that provides many of the required functionalities and
performance. The details of how tasks are mapped or executed on to resources is
abstracted from the \ext user. The user is only required to provide details of
the workload and identify the resource(s) that are accessible. This design
choice thus acknowledges the separation of concerns: workload description from
its execution on HPC systems.

\subsubsection{Configuration files} Composing the advanced sampling algorithms
discussed in the previous sections using the components provided by a particular
middleware stack, requires specific knowledge of the middleware itself. The \ext
framework bypasses this complexity by adding one more level of abstraction. It
provides ready-to-use scripts in accordance with advanced sampling methods
discussed in this paper that are configurable via configuration files. In these
configuration files, the user is exposed to only application-level, meaningful
parameters that can be modified as necessary.

\begin{figure}
\centering
\includegraphics[width=0.45\textwidth]{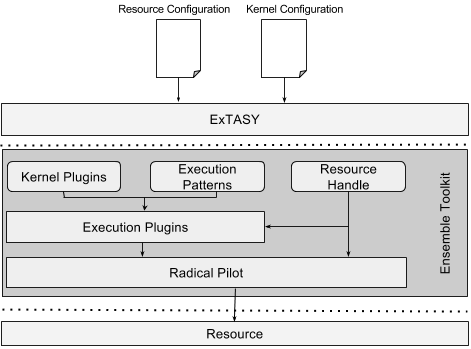}
\caption{Design of the \ext framework using \enmt as middleware. The \ext
  framework provides ready-to-use scripts created using components provided by
  \enmt. Parameters related to the resource and workload are exposed via
  configuration files, which alone are the files that users interact
  with. Within \enmt, the workload is converted into executable units by the
  execution plugins and submitted to the resource using \rp. }
\label{fig:ext_design}
\end{figure}

\subsection{Implementation}

\subsubsection{Ensemble Toolkit}

As mentioned previously, ExTASY requires a middleware for resource and execution
management. We chose to use \enmt~\cite{entk} as the middleware component as it
provides several relevant features such as the ability to support MPI tasks,
dynamic resource management -- one type of which is to be able to execute more
tasks than the resources available, support for heterogeneous HPC systems and
strong and weak scalability guarantees. \enmt has been tested upto O(1,000)
tasks with short and long term plans to support O(10,000) and O(100,000)
tasks~\cite{entk}. \enmt is in turn based upon the pilot abstraction (and the
\rp~\cite{review_radicalpilot} implementation of the pilot abstraction) to provide much of the
flexible and scalable resource management capabilities.

\enmt exposes three components to the user that can be used to express many
applications: Kernel Plugins, Execution Patterns, and Resource Handle. Scripts
that are part of \ext framework use these components to describe the application
logic.

\subsubsection{Configuration files} The application logic is expressed via
components of \enmt. The resource and the workload specifications are exposed
via configuration files. The \ext framework has two types of configuration
files: (i) resource configuration, which consist of details of the resource
where the application will to be executed such as the resource name, the
runtime, and the username and account details used to access the resource, and
(ii) kernel configuration, which defines the workload parameters such as the
location of input files for the Molecular Dynamcis simulation and analysis
tools, parameters for the tools, and workflow parameters such as the number of
simulations.

\section{Applications}

\begin{figure}
\centering
\includegraphics[scale=0.38]{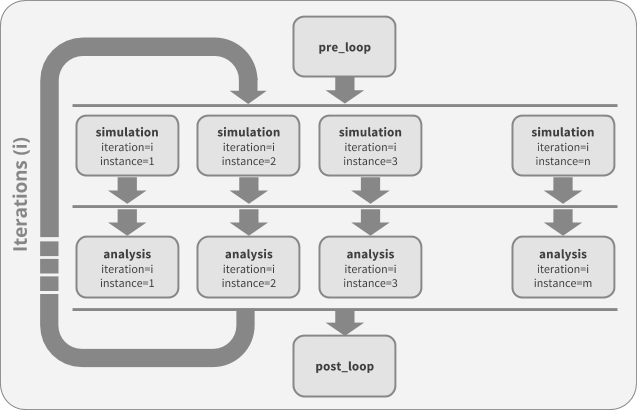}
\caption{The SAL pattern common to both sampling algorithms. The crux of the pattern is an iteration over 2 stages: simulation and analysis, where the number of simulation and analysis instances can be different. The pattern may also contain pre- and post- processing stages.}
\label{fig:sal}
\end{figure}

We illustrate the capabilities of the ExTASY approach via two exemplar applications.
The two different advanced sampling algorithms implemented with
ExTASY are the Diffusion Map-directed-MD (DM-d-MD) and CoCo-MD techniques. Both these algorithms have a common execution pattern: an ensemble of simulation tasks followed by an analysis stage, performed for multiple iterations following the pattern shown in Figure~\ref{fig:sal}.

In the case of the DM-d-MD algorithm, the simulation stage consists of Gromacs  and LSDMap in the analysis stage. Whereas, in the CoCo algorithm, the simulation stage consists Gromacs runs and trajectory conversions and analysis consists of CoCo. The individual simulation or analysis tools might differ depending on the algorithm chosen but the overall pattern is observed to be the same.

\subsection{Diffusion Map-directed-MD}

The Diffusion Map-directed-MD (DM-d-MD) technique \cite{Preto2014} improves
the efficiency of computational resources by choosing which replicas of the
protein are used to run MD. When replicas are too close to each other, the MD
trajectories will be similar. The information gain from simulating MD with
close replicas is small. Part of the replicas which are too close to each other
are deleted. To hold the total number of replicas constant, replicas which are
too far apart from each other are duplicated. In DM-d-MD, a non-linear
dimensionality reduction technique, the locally scaled diffusion map (LSDMap) \cite{Rohrdanz2011} is used to calculate the distance between different
replicas. The deletion or duplication of replicas would destroy the correct
sampling of the protein. By changing the weights of individual replicas in the
reweighting step, the correct sampling of the protein is obtained.

The DM-d-MD technique requires only the protein starting structure. No
additional information about the protein is necessary. The user can fine tune
the sampling mainly by varying the total number of replicas and the way how the
local scale in LSDMap is calculated. At the begin on the method, the replicas
are generated from the protein starting structure. After the MD step, the
LSDMap is calculated. LSDMap requires only the final structure for each replica
from the MD step. Based on the LSDMap results new replicas for the next
iteration of DM-d-MD are chosen from the current replicas. The reweighting
ensures that the 

It was shown that DM-d-MD technique is, at least, one order of magnitude faster
compared to plain MD \cite{Preto2014}. This comparison was done for alanine
dipeptide and a 12-aminoacid model system, Ala12.

\subsection{The CoCo-MD workflow}

The CoCo (\textbf{Co}mplementary \textbf{Co}ordinates) technique~\cite{bib5:CoCo}
was designed originally as a method to enhance the diversity of ensembles of
molecular structures of the type produced by NMR structure determination. The
method involves the use of PCA \cite{jollife:PCA,bib3:Sherer,bib4:Wlodek} in
Cartesian space to map the distribution of the ensemble in a low (typically 2-4
dimensional) space, and then the identification of un-sampled regions. CoCo
generates new conformations for the molecule that would correspond to these
un-sampled regions. The number of new structures generated is under the user’s
control – the algorithm divides the space into bins at a chosen resolution,
marks bins as sampled or not, first returns a structure corresponding to the
centre of the un-sampled bin furthest from any sampled one, marks this bin as
now sampled, and iterates as many times as desired. 

In the CoCo-MD workflow, an ensemble of structures from MD simulations are
analysed using the CoCo method; new conformations become the start points for a
new round of MD simulations. The latest MD data is added to the previous set,
and CoCo repeated. The method is agglomerative -– all MD data generated so far
is used for each analysis; but also adaptive –- a fresh PCA is performed each
time. Applied to simulations of the alanine pentapeptide, the CoCo-MD workflow
is able to reduce mean first passage times from the extended state to other
local minimum states by factors of ten or greater compared to conventional
simulations \cite{coco-unpublished}.

\section{Performance Evaluation}\label{sec: perf_eval}

\subsection{Experiment setup}

\subsubsection{Physical system}

The 39-residue mixed $\alpha$/$\beta$ protein NTL9(1-39) (pdb code 2HBA, 14,100
atoms including water) is chosen as the physical system for our experiments.
NTL9 has an experimentally measured folding time of around 1.5 ms
\cite{Horng20031261}, and its folding process has been extensively studied by
experiment and all-atom MD simulations, both by means of the Folding@Home
distributed computing platform coupled with MSM analysis \cite{Voelz2010}, and
by Anton supercomputer \cite{Lindorff-Larsen2011}.

The relatively small size of NTL9, and the existence of previous MD simulation
results over long timescales, make this protein an ideal candidate for testing
and benchmarking our approach.  Albeit small, NTL9 is much larger than a simple
peptide, and exhibits a folding process with two competing routes
\cite{Lindorff-Larsen2011}, thus presenting a non-trivial test for adaptive
sampling.

\subsubsection{HPC systems used}

One of the requirements of ExTASY as that it should be interoperable, so we
have used several different HPC systems for our experiments, and characterised
the performance of ExTASY on each.

{\bf Stampede} is a Dell Linux cluster located at the Texas Advanced Computing Center,
and is part of the Extreme Science and Engineering Discovery Environment
(XSEDE).  It consists of 6400 compute nodes, each with 2 Intel Xeon `Sandy
Bridge' processors, for a total of 16 CPU cores per node, as well as an Intel
Xeon Phi co-processor (not used in our experiments).  Stampede uses the SLURM
batch scheduler for job submission.

{\bf ARCHER} is a Cray XC30 supercomputer hosted by EPCC, and operated on behalf of
the Engineering and Physical Sciences Research Council (EPSRC) and the Natural
Enviroment Research Council (NERC).  It has 4920 compute nodes, each with 2
Intel Xeon `Ivy Bridge' processors, giving 24 cores per node.  ARCHER uses the
Portable Batch System (PBS) for job submission.

{\bf Blue Waters} is a Cray XE6/XK7 operated by the National Centre for Supercomputing
Applications on behalf of the National Science Foundation and the University of
Illinois.  The XE6 partition used in this work consists of 22640 compute nodes
with 2 AMD `Interlagos' processors, giving 32 cores per node.  Blue Waters uses
the TORQUE/Moab workload manager for job submission.

\subsection{Evaluation of individual components}

Since the performance of the entire workflow depends on the performance of each of the component parts, we investigate the scaling of both the simulation code (Gromacs) and the analysis tools in isolation on each of the three target platforms, using the NTL9 system used for the full sampling workflows.

\subsubsection{Simulation tools}

\begin{figure}[!t]
\centering
\includegraphics[width=0.48\textwidth]{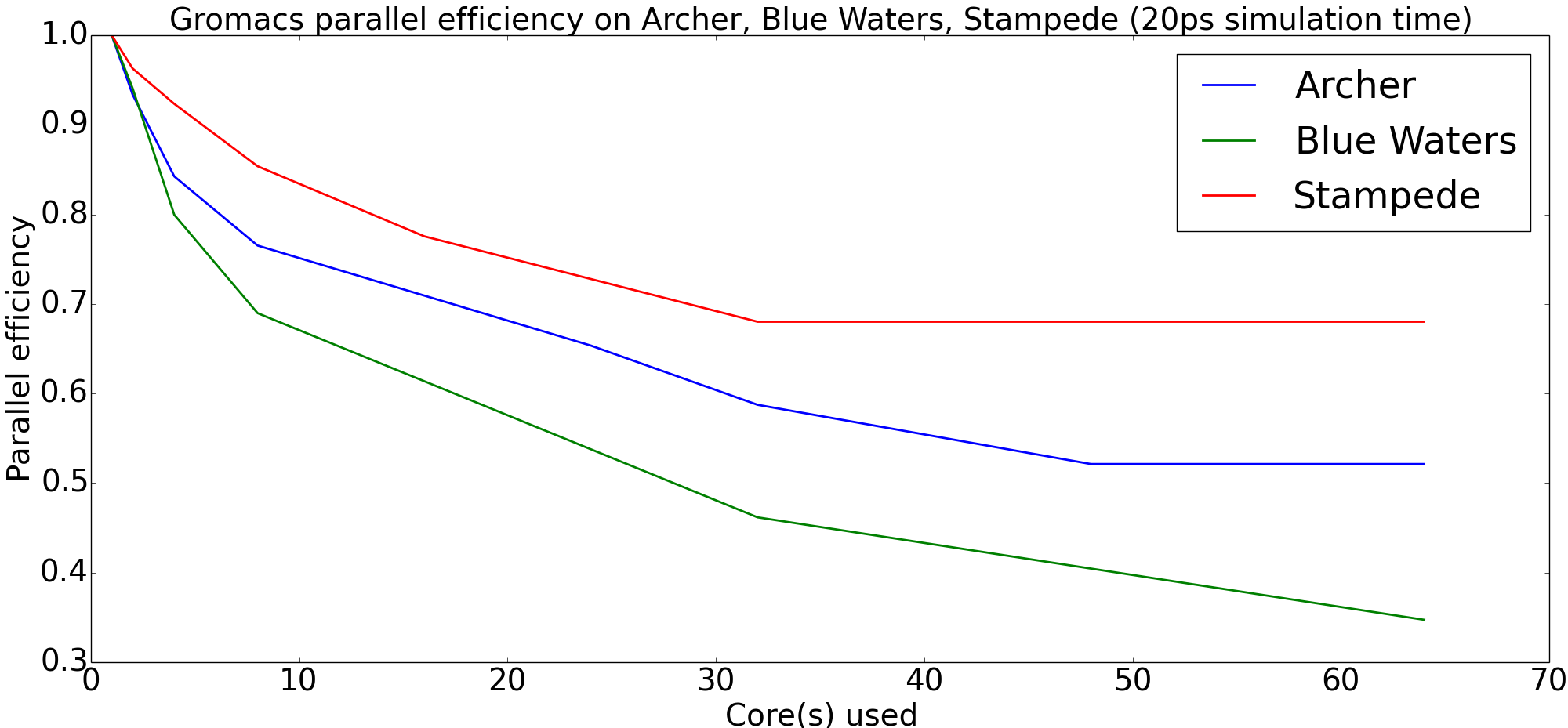}
\caption{Gromacs parallel efficiency on ARCHER, Blue Waters and Stampede. A single 20ps gromacs simulation of the NTL9 system is performed using various core counts on the three machines and the execution time is measured.}
\label{fig:gmx-eff}
\end{figure}

The parallel efficiency of Gromacs with respect to a single core on each machine is shown in Figure
\ref{fig:gmx-eff}.
While efficincies of 69\% (ARCHER, 24 cores), 78\%
(Stampede, 16 cores) and 46\% (Blue Waters, 32 cores) suggest that while the
scaling for such a relatively small simulation is not ideal, using a single node
per simulation is a good use of the available hardware.  Beyond a single node,
the efficiency drops off so although multiple node simulation tasks are
supported by Ensemble Toolkit they are not useful for this benchmark case.

\subsubsection{Analysis tools}
\label{sec:tools}

Due to the nature of the two workflows, there are many parallel simulation
tasks, but only a single analysis task.  Therefore, the analysis task may be
configured to run on as many cores as are available to the simulations.  Both
CoCo and LSDMap are parallelised using MPI, and consist
of parts which are independent, e.g., reading of
  trajectory files in CoCo, and involve communication e.g. diagonalisation of
  the covariance matrix in CoCo and the diffusion matrix in LSDMap, so the parallel scaling
is expected to be sub-linear.
The performance of CoCo is also strongly dependent on I/O since it reads the
entire trajectory file rather than just the final configurations like LSDMap.

\begin{figure}[!t]
\centering
\includegraphics[width=0.48\textwidth]{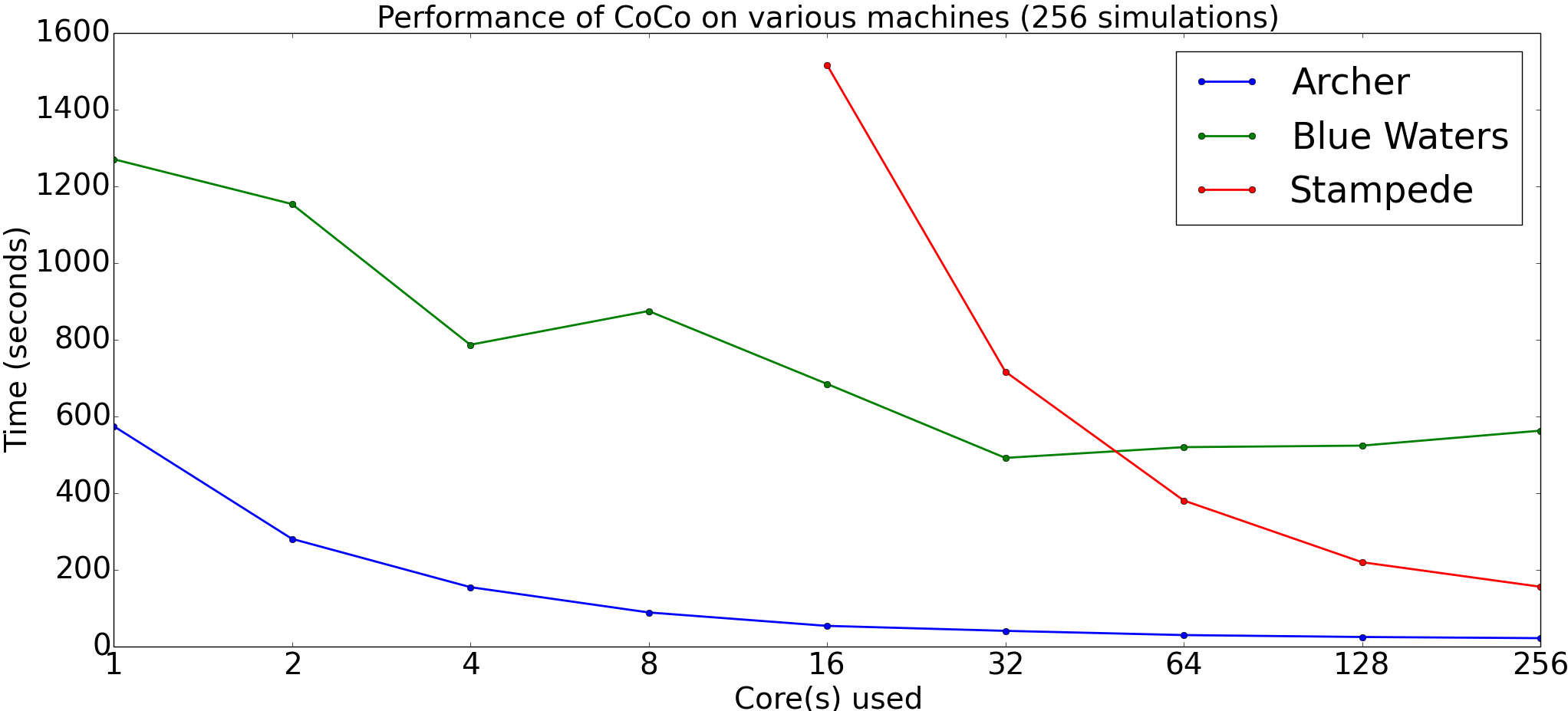}
\caption{Strong scaling of CoCo analysis tool on ARCHER, Blue Waters and Stampede. A total of 256 simulations are analyzed using various core counts on the three machines and execution time is measured.}
\label{fig:coco-perf}
\end{figure}

\begin{figure}[!t]
\centering
\includegraphics[width=0.48\textwidth]{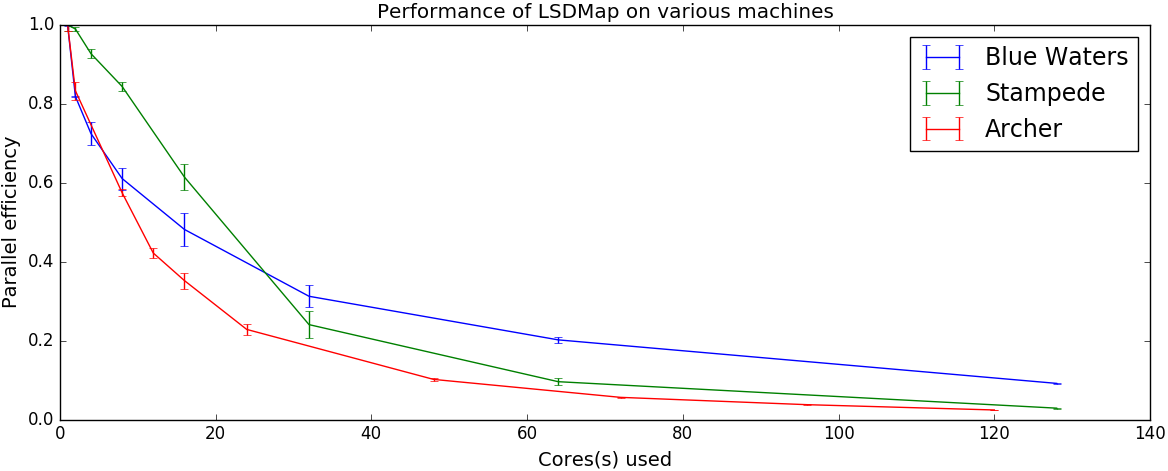}
\caption{Parallel efficiency of LSDMap on ARCHER, Blue Waters and Stampede. A total of 2449 structures are analyzed using various core counts on the three machines and execution time is measured.}
\label{fig:lsdmap-perf}
\end{figure}

Figures \ref{fig:coco-perf}
show the strong scaling of CoCo for a fixed input of 256 simulations.  We
see that CoCo is able to scale to at least 256 cores on ARCHER and Blue Waters,
and to around 32 cores on Stampede, thus
for our following experiments we configure the workflow to run CoCo with as many
cores as there are input trajectories.
LSDMap (Figure \ref{fig:lsdmap-perf}) however, does not scale efficiently much beyond a single node on each machine, even with over 2000 input structures.  Nevertheless, we run LSDMap on as many cores as are available, even though it cannot use them fully.  Due to the structure of the workflow, those cores would otherwise sit idle during the analysis phase.

\subsection{Evaluation of ExTASY}

\subsubsection{Characterization of overheads}
\label{sec:overhead}

In addition to the necessity of characterizing performance overhead introduced
by a new system approach and implementation, in order to instill confidence in
potential users of \ext it is important to measure the overheads imposed by
\ext.  The objective is to discern the contributions from the different aspects
of the \ext framework as opposed to MD and analysis components.  The time taken
to create, launch and terminate all of the simulations or the instantiate
\ext framework itself on the resource are examples of the former overhead.
All of the experiments use \enmt version 0.3.14.

\begin{figure}[!t]
\centering
\includegraphics[width=0.48\textwidth, height=0.22\textwidth]{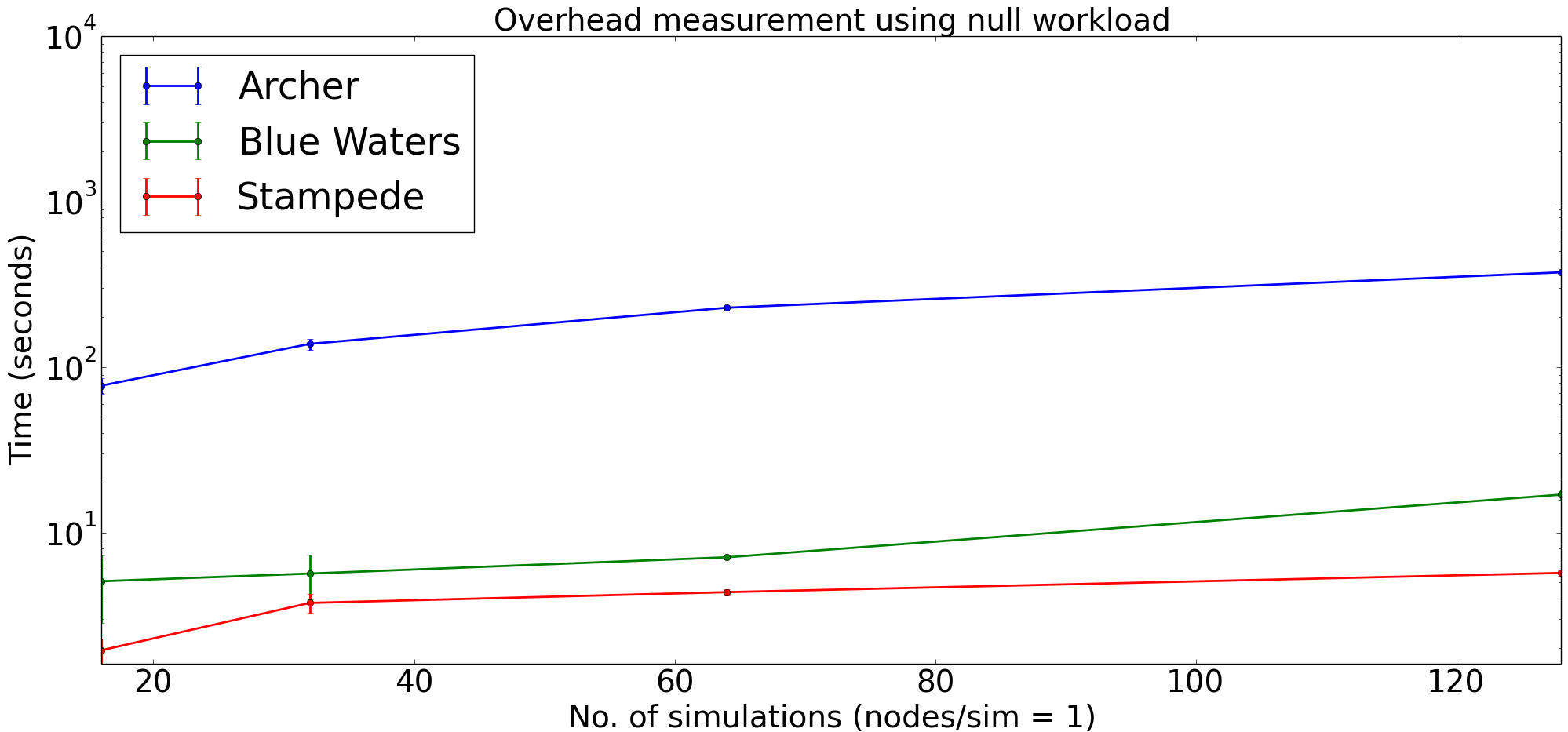}
\caption{Measurement of overhead from \ext on ARCHER, Blue Waters and Stampede. Using only the simulation stage of the SAL pattern, overhead from \ext is measured on the three machines at various core counts. }
\label{fig:overhead}
\end{figure}

We ran a single iteration of the workflow with null
workloads, i.e., where the task did no work (\texttt{/bin/sleep 0}), but
otherwise was configured as a ``normal'' simulation task, launched using MPI and
taking a whole node on each of the machines. The number of tasks ranged from 16
to 128, and they were all run concurrently.
Figure \ref{fig:overhead} shows that the overheads on Stampede and
Blue Waters are relatively small, growing from $<$5s to around 15s for 128
tasks. On ARCHER the overheads are much larger (70-350s), which after further
investigation is due to the \texttt{aprun} job launcher, which performs
poorly when multiple concurrent tasks are launched.

\subsubsection{Strong scaling test}
\label{sec:strong-scaling}

To test the strong scaling of the ExTASY workflows, we fix the number of
executing instances (independent MD simulations) to 128 and vary the total
number of CPU cores used to execute the workflow.  The largest 
experiments use enough CPU cores that all of the MD simulations execute
concurrently.  For example, on ARCHER, 128 instances, each executing on a single
node (24 cores) gives a maximum of 3072 cores. 

Since the MD simulations are independent and may be executed concurrently by
ExTASY, we expect that as the number of cores is increased, the simulation time
should decrease proportionally.  The time spent in the analysis part is expected
to be constant since the total amount of MD data is constant, and despite the
parallelisation of the analysis tools as can be seen from Figure 5, the time to
completion plateaus at fairly low core counts.

\begin{figure}[!t]
\centering
\includegraphics[width=0.48\textwidth]{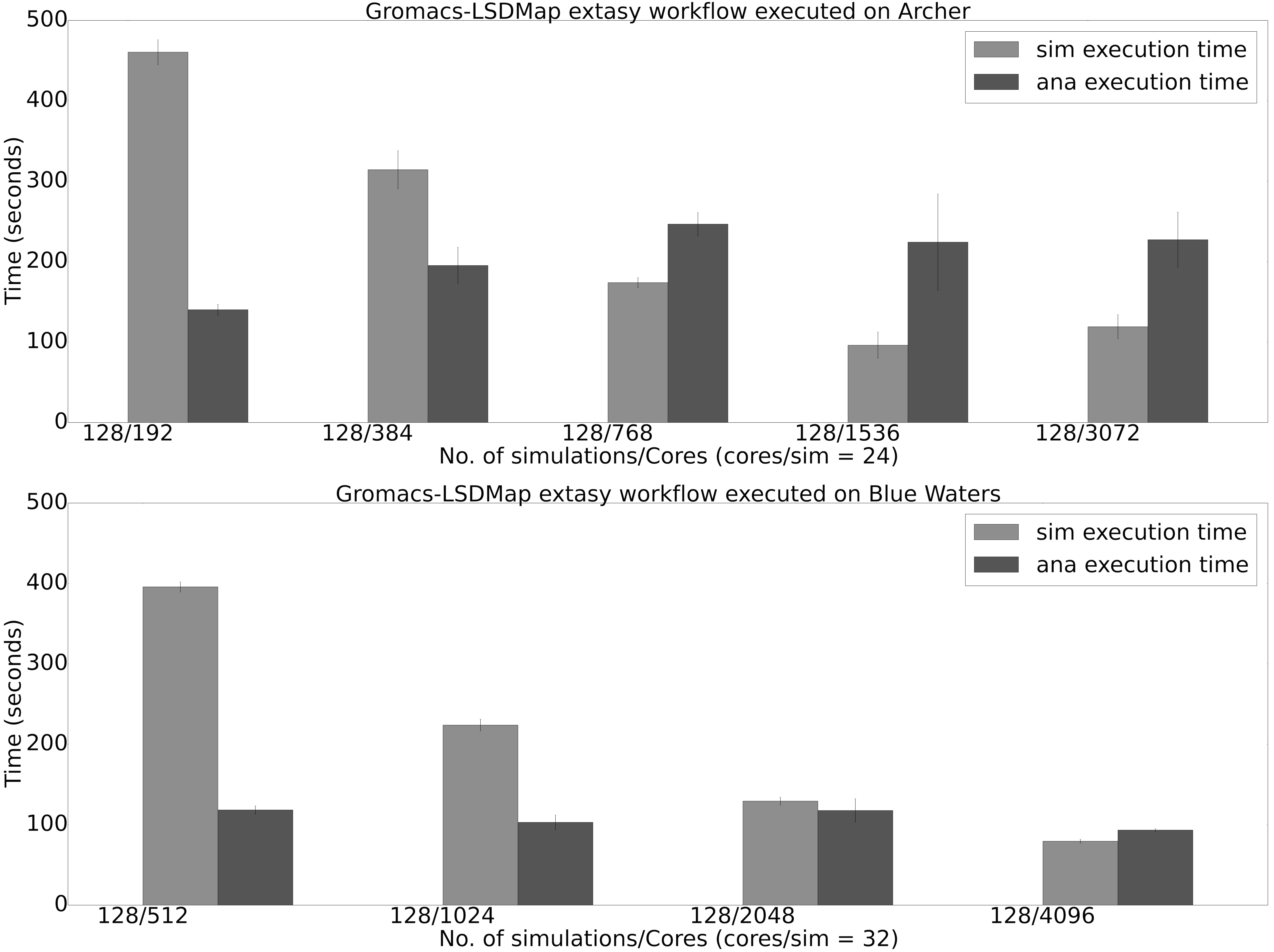}
\caption{Strong scaling of DM-d-MD workflow on ARCHER
  (\textbf{top}) and Blue Waters (\textbf{bottom}). The number of simulations is
  held constant at 128, number of cores per simulation at 24 on ARCHER and 32 on
  Bluewaters. The total number of cores used is varied with a constant workload,
  hence measuring strong scaling performance of the framework.
}
\label{fig:grlsd-on-archer-bw-strong}
\end{figure}

Figure~\ref{fig:grlsd-on-archer-bw-strong} (\textbf{bottom}) shows the results of this experiment for the
DM-d-MD workflow executed on Blue Waters.  The simulation time decreases from
395.7s on 512 cores to 79.52s on 4096 cores, a speedup of 4.97x with 8x as many
cores -- yielding a scaling efficiency of 62\%.  The analysis time is
essentially constant at around 100s, as expected.  The loss of scaling
efficiency for the simulation part comes from two sources.  Firstly, there is
the fixed overhead discussed in Section~\ref{sec:overhead} associated with the
execution of 128 concurrent tasks, which is approximately 15s.  Secondly, the
actual computation which occurs within each task take longer when more
simulations are run concurrently, due to the fact that they all write to the
same shared filesystem.  For example, when 16 instances are run concurrently on
512 cores, the MD simulations take an average of 45.6s each.  When all 128
instances are run concurrently, each takes 49.0s, or 3.4s slower.  If these
effects are removed, the effective scaling efficiency on 4096 cores rises to
77\%.

Similar results are obtained on ARCHER
(Figure~\ref{fig:grlsd-on-archer-bw-strong}, \textbf{top}), although the scaling of the
simulation part tails off on 3072 cores, and the LSDMap analysis takes somewhat
longer, with higher variability than Blue Waters.  Both of these are due to the
fact that both the MD and analysis involve significant I/O, and it is known that
opening many small files concurrently is slow as the metadata servers of the
parallel Lustre filesystem become a bottleneck \cite{ARCHER-IO}.

\begin{figure}[!t]
\centering
\includegraphics[width=0.48\textwidth]{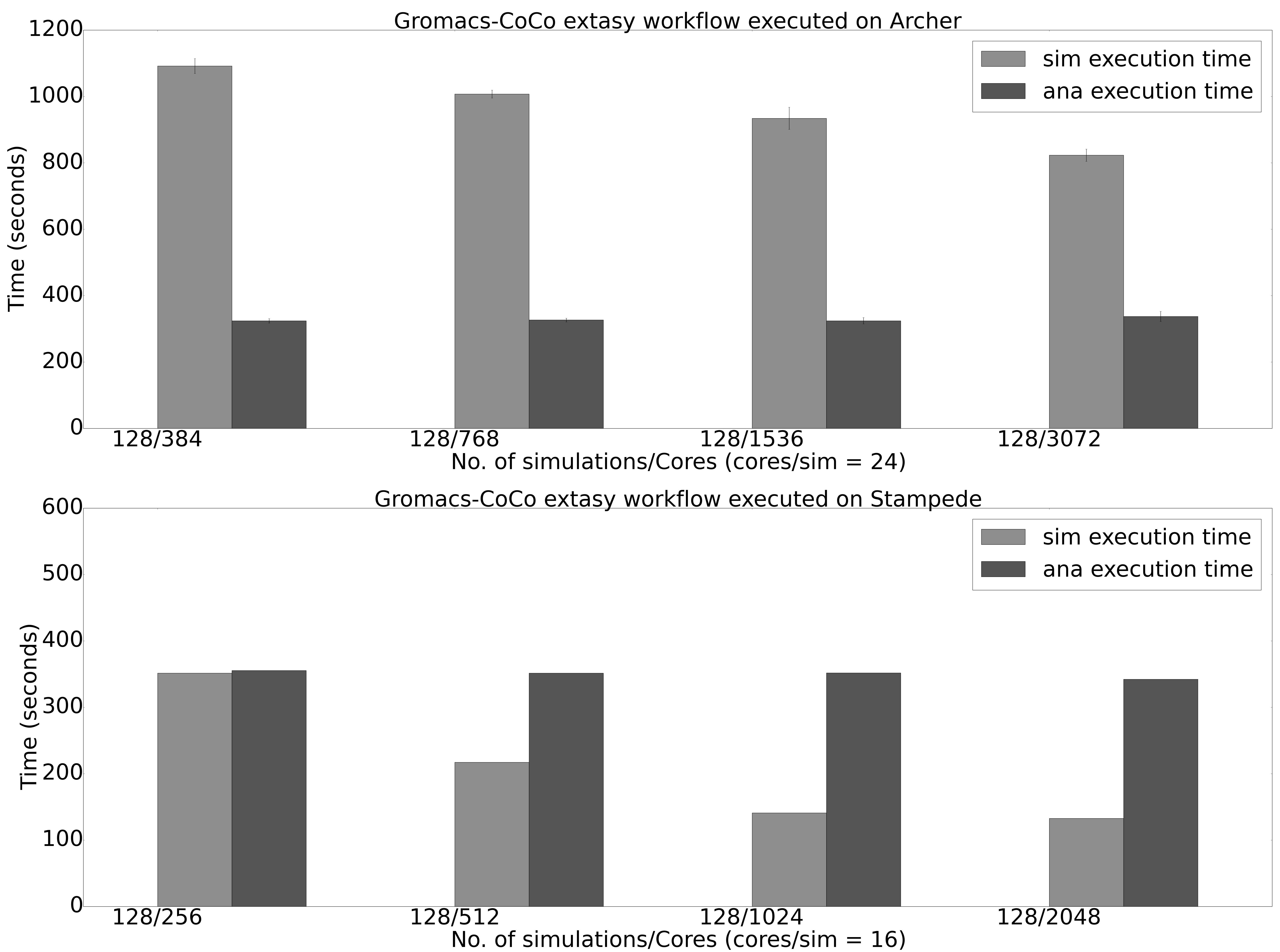}
\caption{Strong scaling of CoCo-MD workflow on ARCHER (\textbf{top}) and
  Stampede (\textbf{bottom}). The number of simulations is held constant at 128,
  number of cores per simulation at 24 on ARCHER and 16 on Stampede. The total
  number of cores used is varied with a constant workload, hence measuring
  strong scale performance of the framework.
}
\label{fig:gmxcoco-on-archer-stampede-strong}
\end{figure}

CoCo-MD on Stampede (Figure \ref{fig:gmxcoco-on-archer-stampede-strong}, \textbf{bottom}) has similar
strong scaling for the simulation part as DM-d-MD.  The simulation time
decreases from 363s on 256 cores to 83.7s on 2048 cores -- a speedup of 4.3x for
a 8-fold increase in the number of cores (54\% efficiency).  However, the
analysis time (CoCo) does not scale due to the fact that the
parallelisation in CoCo is limited to the number of input trajectories, which is
128 in this case, even if more cores are available.

The CoCo-MD workflow on ARCHER
(Figure\ref{fig:gmxcoco-on-archer-stampede-strong}, \textbf{top}) does not show
as good scaling as DM-d-MD, or CoCo-MD on the other platforms.  The reason for
this lies in the fact that after the actual molecular dynamics calculation, a
`trajectory conversion' step is required to prepare the data for CoCo.  This
step only takes a fraction of a second to execute, but there is a very large
overhead caused by \texttt{aprun}, which allocates resources to and launches
each individual task.  This does not occur on Blue Waters, which uses the
ORTE\cite{rp-orte} implementation of \rp that is not yet default on ARCHER.

\subsubsection{Weak scaling test}

To investigate the weak scaling properties of \ext, we fix the ratio of number
of instances to CPU cores, and vary the number of instances with the constraint
that all simulations can execute concurrently.  For example, on ARCHER 16
instances are executed on one node each (24 cores) giving a total of 384 cores,
and the number of instances is increased to 128, i.e., the number of cores is
3072.

Since all simulations run concurrently, and the length of each simulation does
not change, we expect the simulation time to be a constant.  However, the
analysis part will increase since the performance of the analysis tools is a
function both of the input data size (depending on the number of instances), as
well as the number of cores available, and even though the number of cores is proportional
to the data size, the amount of work grows faster than linearly with the data size.

\begin{figure}[!t]
\centering
\includegraphics[width=0.48\textwidth]{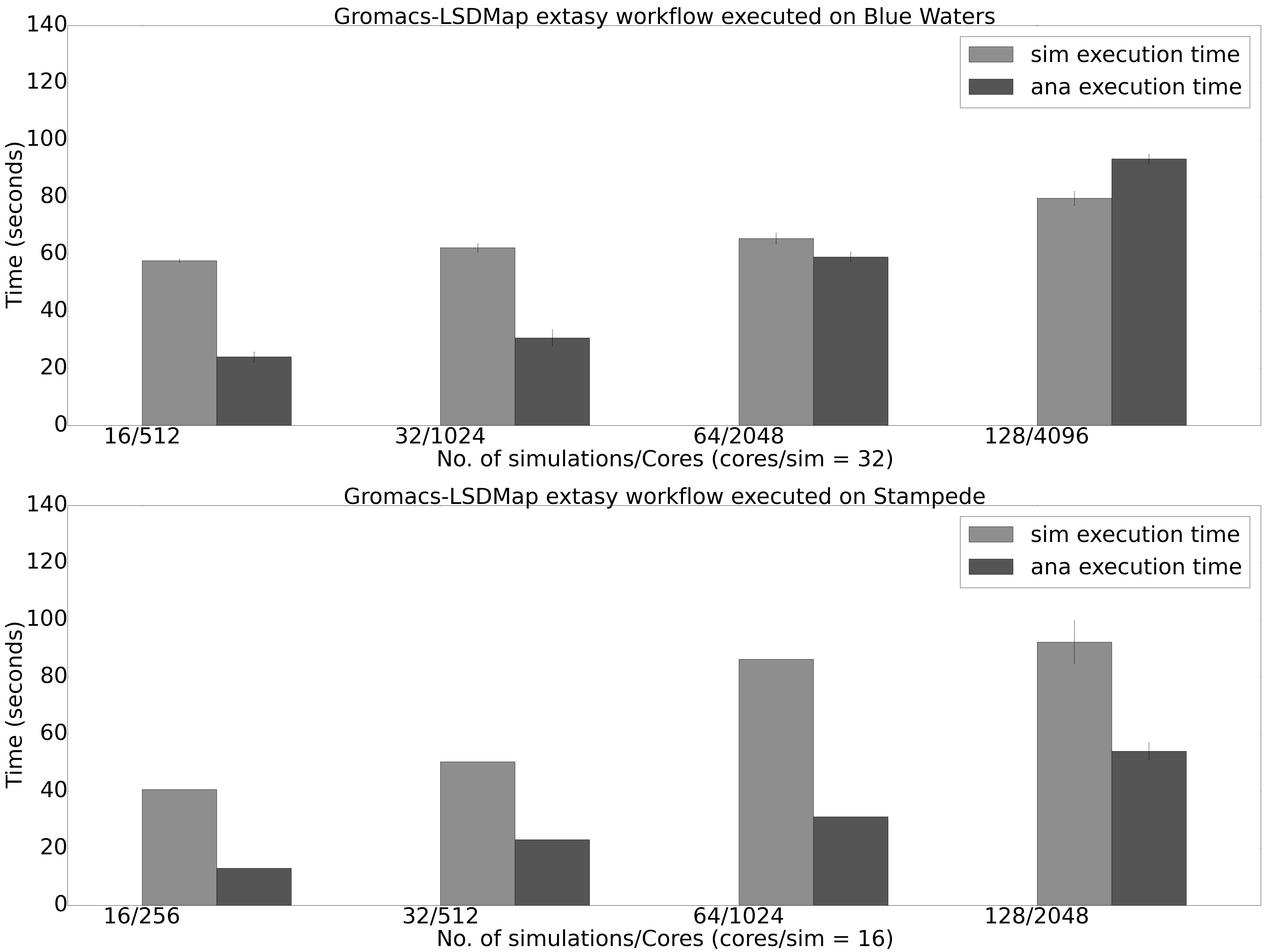}
\caption{Weak scaling of DM-d-MD workflow on Blue Waters (\textbf{top}) and Stampede (\textbf{bottom}). The number of cores per simulation is held constant at 32 on Blue Waters and 16 on Stampede. The total number of simulations is varied from 16-128 and the cores used are increased proportionally. By keeping the ratio of the workload to the number of resources constant, we observe the weak scaling performance of the framework.
}
\label{fig:grlsd-on-bw-stampede-weak}
\end{figure}

For the DM-d-MD workflow on Blue Waters (Figure~\ref{fig:grlsd-on-bw-stampede-weak}, \textbf{top}) we
observe a small increase of 21.8s in the simulation part as we scale from 512 to
4096 cores.  Similar to the strong scaling results, this is combination of the
overhead due to the increased number of tasks with a slowdown of the individual
tasks themselves.  The analysis is found to increase sub-linearly.  As discussed
in section \ref{sec:tools} the LSDMap computation consists of parts which are
both linear and quadratic in the size of the input data.  Combined with an
increasing number of cores available for the tool, sub-linear scaling is the
result.  Similar behaviour is observed on Stampede
(Figure~\ref{fig:grlsd-on-bw-stampede-weak}, \textbf{bottom}), although with a different weighting
of the simulation and analysis part, reflecting the fact that the performance of
each kernel depends on how well optimised the application binary is on the
execution platform.

\begin{figure}[!t]
\centering
\includegraphics[width=0.48\textwidth]{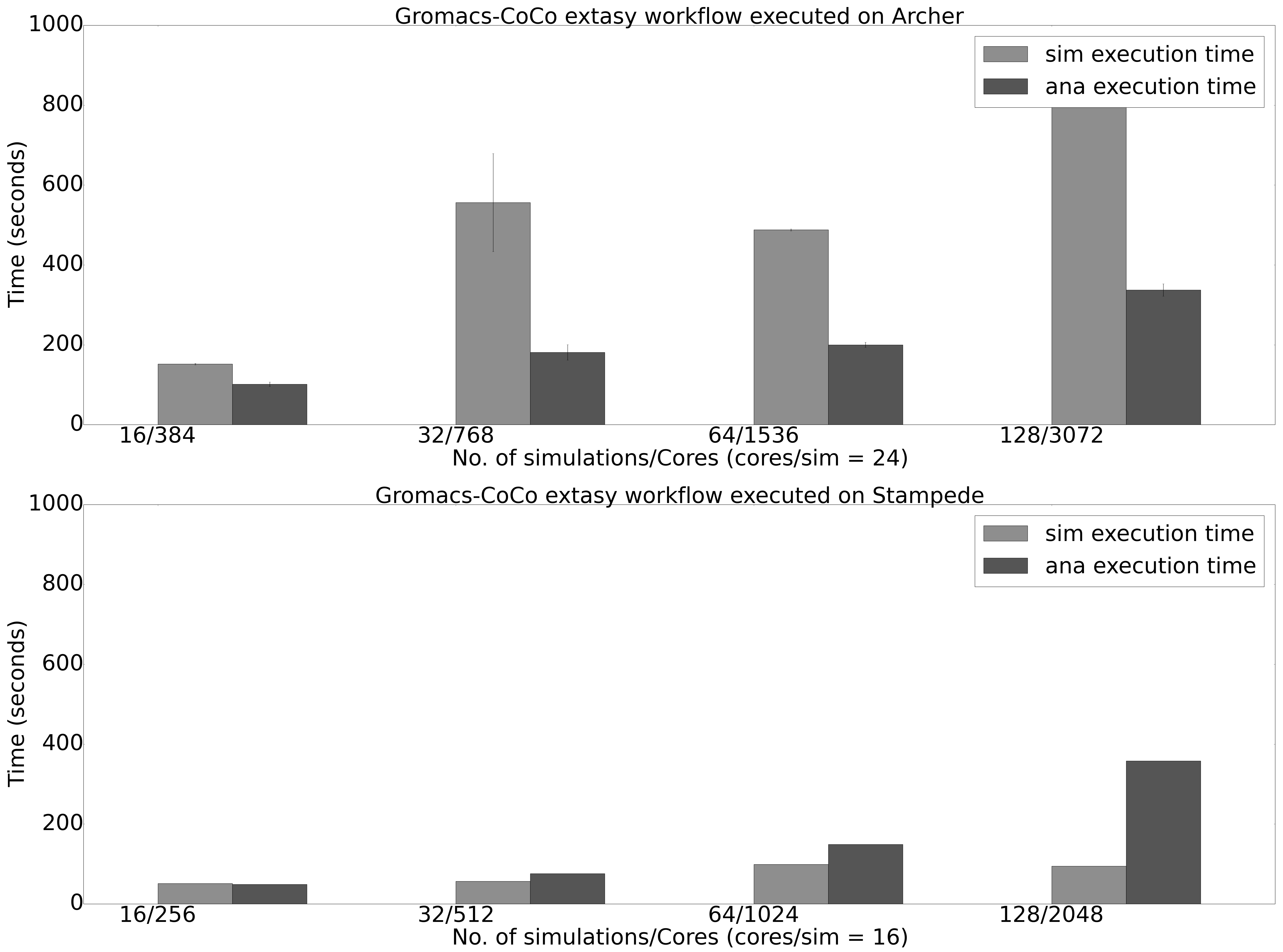}
\caption{Weak scaling of CoCo-MD workflow on ARCHER (\textbf{top}) and Stampede (\textbf{bottom}). The number of cores per simulation is held constant at 24 on ARCHER and 16 on Stampede. The total number of instances is varied from 16-128 and the cores used are increased proportionally. By keeping the ratio of the workload to the number of resources constant, we observe the weak scaling performance of the framework.
}
\label{fig:gmxcoco-on-archer-stampede-weak}
\end{figure}

The weak-scaling of CoCo-MD on ARCHER (Figure~\ref{fig:gmxcoco-on-archer-stampede-weak}, \textbf{top})
shows very clearly the aprun bottleneck discussion in Section
\ref{sec:strong-scaling}, and the effect increases as the number of concurrent
tasks grows.  However, the analysis part scales better than linearly, which is
to be expected since CoCo consists of parts which weak scale ideally
(independent operations per trajectory file) and parts such as the construction
and diagonalisation of the covariance matrix which grow as the data size
squared or more.

On Stampede, the weak scaling of the simulation part of the CoCo-MD workflow (Figure \ref{fig:gmxcoco-on-archer-stampede-weak}, \textbf{bottom}) is much better than ARCHER.  The simulation time grows only by around 50s compared to over 700s on ARCHER over the range of cores that we tested.  CoCo scales almost identically to ARCHER.

\subsubsection{Effect of larger ensembles}

\begin{figure}[!t]
\centering
\includegraphics[width=0.48\textwidth]{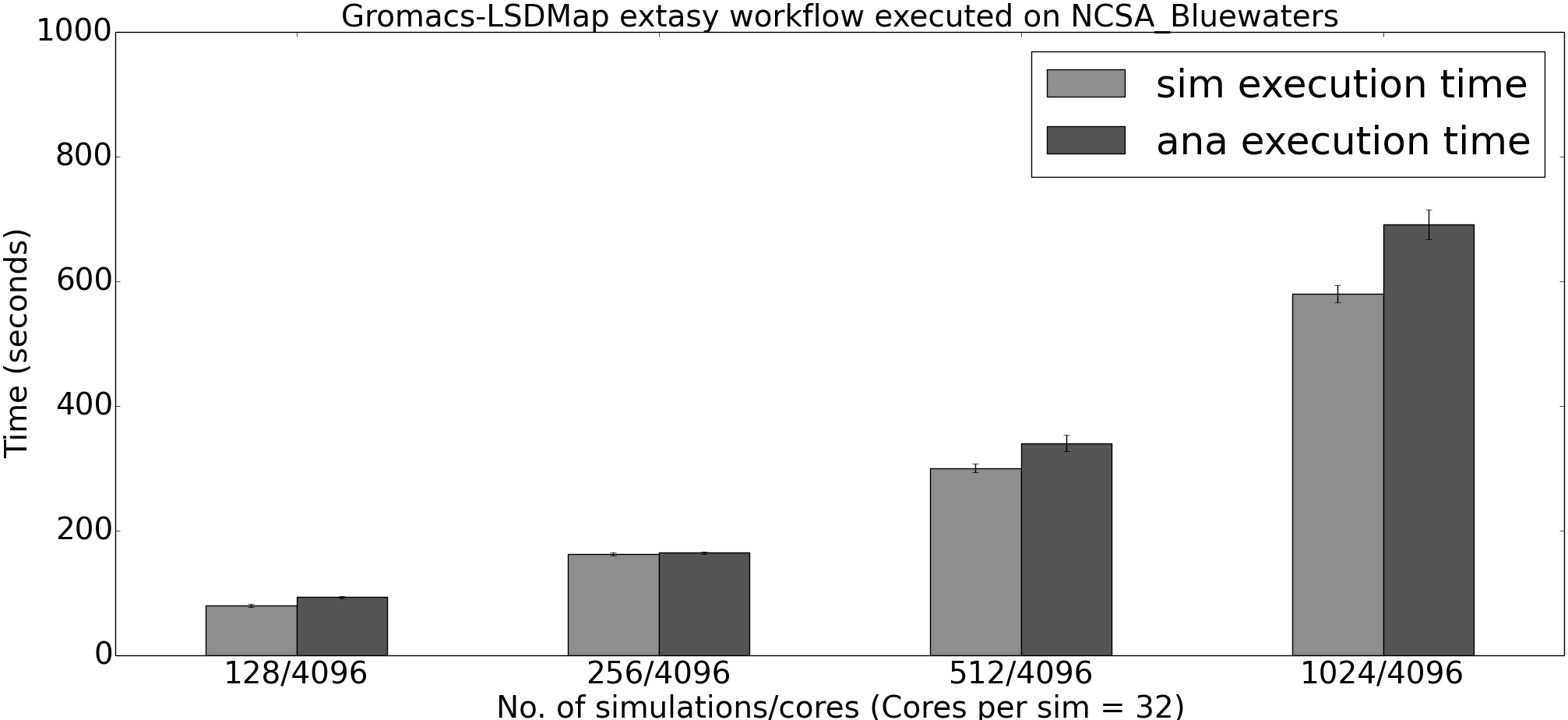}
\caption{DM-d-MD workflow on Stampede. Workload is increased from 128 instances to 1024,
  keeping the number of
  cores constant at 4096. Within the overheads, the increase in execution time
  is in proportion with increase in the workload.}
\label{fig:grlsd-para}
\end{figure}

To distinguish the effects caused by strong scaling (increasing parallelism with
a fixed amount of work) and weak scaling (increasing parallelism proportionally
to the amount of work), we also measured the effect of increasing the amount of
work with a fixed number of compute cores available.  Figure
\ref{fig:grlsd-para} shows the results for the DM-d-MD workflow running on
Blue Waters as we vary the number of MD instances from 128 to 1024, keeping the
total number of cores available at 4096.  Since each task runs on a single node
(32 cores per instance), only 128 simulation tasks can run concurrently.
Ideally, we would expect that the simulation time should increase linearly with
the number of instances.  In practice, we see that the time taken grows by only
7.4x as the number of instances increases from 128 and 1024 i.e., a factor
of 8.  This is due to the fact that some of the overheads related to managing
the tasks that occur before or after execution in the 128 task case are one-time
overheads, i.e., those overheads are hidden as they are done concurrently (in
the \rp Agent) with the execution of the remaining tasks when the number of
instances is greater than 128.  The scaling of the analysis part is
consistent with that discussed in Section~\ref{sec:tools}, that there is a close to linear
scaling since the larger the ensemble size, the more parallelism is available in LSDMap

\subsubsection{Dynamic simulations}

\begin{figure}[!t]
\centering
\includegraphics[width=0.48\textwidth]{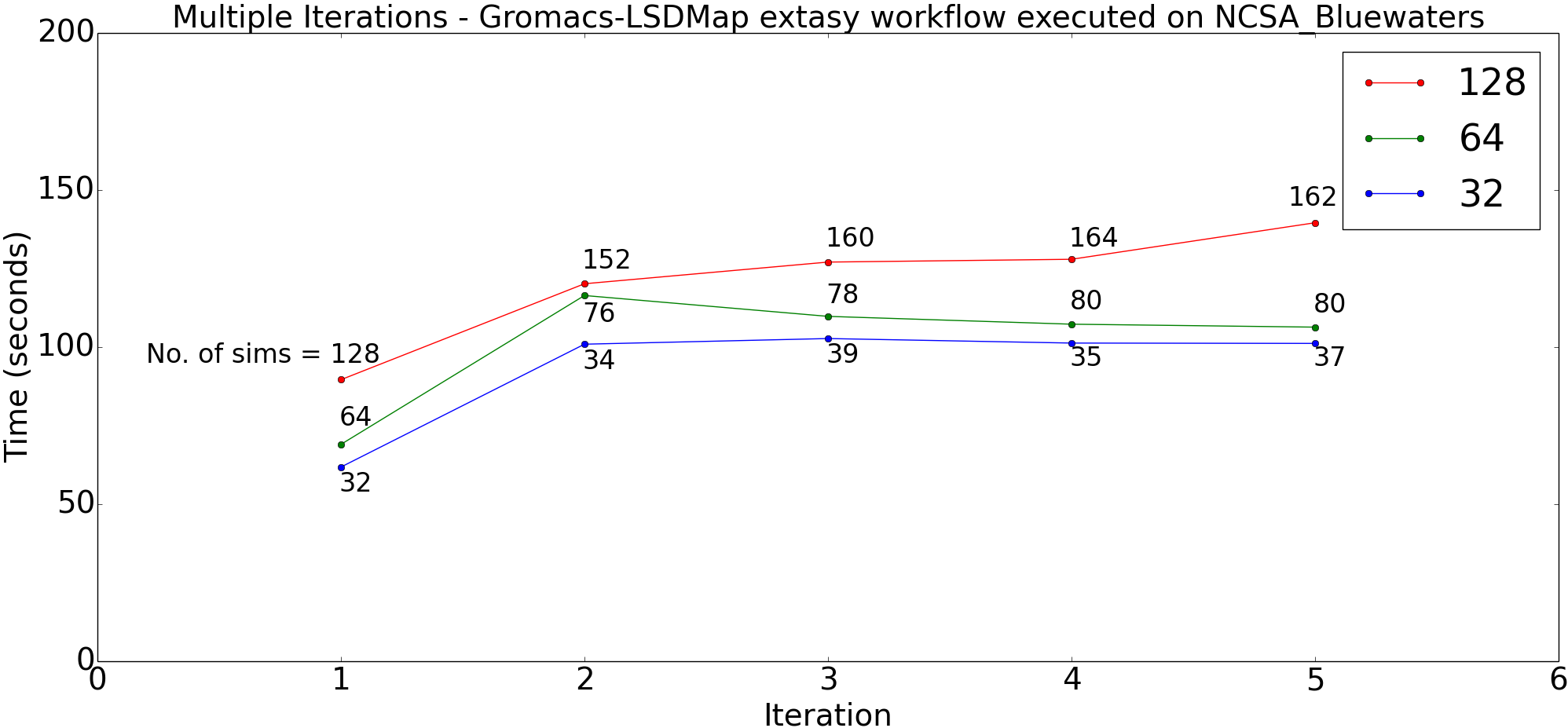}
\caption{Support for dynamic workload in ExTASY: The DM-d-MD algorithm dictates the number of instances at every iteration. The number of instances in each iteration (for a total of 5 iterations) when starting with 32, 64, 128 instances is presented.}
\label{fig:dynamic}
\end{figure}

An important characteristic of the LSDMap and CoCo based workflows is that the
number of instances typically changes after each simulation-analysis
iteration. Thanks to the pilot-abstraction, the \ext framework supports flexible
mapping between the number of concurrent instances and the total number of
cores, while being agnostic of the number of cores per instance.  This
functionality is used by the DM-d-MD workflow, where, depending on the progress
through the conformation space of the system being explored, LSDMap may decide
to spawn more (or less) trajectories for the next iteration of sampling.  Figure
\ref{fig:dynamic} illustrates this capability.  We ran the DM-d-MD workflow on
Blue Waters for three configurations with 32, 64 and 128 initial instances.  We
can see that after an initial growth phase the number of instances seems to
stablise for the remaining iterations, although the difference from the starting
configuration and the number of iterations taken to stabilise is not
algorithmically or systematically predictable.  The flexible resource
utilization capabilities that \ext is built upon prove critical.

\subsection{Summary of Experiments}

We have shown illustrative performance data for two different applications --
CoCo-MD and DM-d-MD -- based on different analysis methods, on three distinct
HPC platforms - ARCHER, Blue Waters and Stampede.  The overall scaling to
O(1000) simulations is clearly demonstated, and we have analysed the scaling
behaviour of the \ext framework itself (overheads) and the individual simulation
and analysis programs which constitute the workflow.

\section{Discussion and Conclusion}

State-of the-art computational biophysics approaches aim to balance three
different requirements: force-fields accuracy, advanced sampling capabilities,
and rigorous and fast data analysis. These points are strongly interconnected.
In particular, it is becoming clear that advanced sampling and data analysis
need to be tightly coupled to work efficiently and accurately, as the
configurational space that has already been sampled needs to be analyzed
on-the-fly to inform on how to proceed with further sampling.  Furthermore, many
advanced sampling algorithms for biomolecular simulations require flexible and
scalable support for multiple simulations.  As no final solution yet exists on
the best strategy for adaptive sampling (and different physical systems may require a
combination of strategies), there is a need to allow combination of different MD
engines with different analysis tools, and can be easily extended or modified by
end-users to build their own workflows, both for development of new strategies,
and for applications to the study of complex biomolecular systems.
 
\ext is designed and implemented to provide a significant step in this
direction.  \ext allows to simulate many parallel MD trajectories (by means of
standard MD engines), to extract long timescale information from the
trajectories (by means of different dimensionality reduction methods), and to
use the information extracted from the data for adaptively improving sampling.
\ext has been utilized by different MD engines and analysis algorithms, with
only pre-defined and localized changes.

In Section~\ref{sec:soft_arch}, we formally identified the functional,
performance and usability requirements to support the coupling of MD simulations
with advanced sampling algorithms. We then presented the design and
implementation of \ext, an \emph{extensible}, \emph{portable} and
\emph{scalable} Python framework for building advanced sampling workflow
applications to achieve these requirements.  After establishing accurate
estimates of the overhead of using \ext, Section~\ref{sec: perf_eval} consisted
of experiments designed to validate the design objectives; we performed
experiments that characterized \ext along traditional scaling metrics, but also
investigated \ext beyond single weak and strong scaling performance.  With the
exception of some machine specific reasons,\ext displayed linear scaling for
both strong and weak scaling tests on various machines up to O(1000) simulation
instances on up to O(1000) nodes for both DM-d-MD and CoCo-MD workflows.

In order to keep the footprint of new software small, \ext builds upon
well-defined and understood abstractions and their efficient and interoperable
implementation (\enmt, \rp).  This provides double duty: The core
functionality of \ext can be provided by simple higher level extensions of
complex system software, while allowing it to build upon the performance and
optimization of the underlying system software layers.  This also allows \ext 
to employ good systems engineering practice: well-defined and good base
performance, while being amenable to platform-specific optimizations (e.g.
using ORTE on Blue Waters~\cite{rp-orte}).

The design of \ext to reuse existing capabilities, for extensibility to
different MD codes and sampling algorithms while providing well defined
functionality and performance are essential features to ensure the
sustainability of \ext.  Compared to existing software tools and libraries for
advanced sampling, \ext provides a much more flexible approach that is agnostic
of individual tools and compute platforms, is architected to enable efficient
and scalable performance and has a simple but general user interface.  The \ext
toolkit is freely available from \url{http://www.extasy-project.org}.

The \ext toolkit has been used to deliver two hands-on computational science
training exercises and tutorials to the bio-molecular simulations community with
a focus on advanced sampling. Participants were given the opportunity to utilize
HPC systems in real time for advanced sampling problems of their own. Details of
both events can be found at
\url{http://extasy-project.org/events.html#epccmay2016}. A link to the lessons
and experience from the first workshop can be found at:
\url{https://goo.gl/nMSd27}.

\section{Acknowledgments}

{\footnotesize This work was funded by the NSF SSI Awards (CHE-1265788 and
  CHE-1265929) and EPSRC (EP/K039490/1).  This work used the ARCHER UK National
  Supercomputing Service (http://www.archer.ac.uk).  We acknowledge access to
  XSEDE computational facilities via TG-MCB090174 and Blue Waters via
  NSF-1516469. We gratefully acknowledge the input from various people who have
  helped the development of the ExTASY workflows: everyone else involved the
  ExTASY project, the attendees at ExTASY tutorials and beta testing sessions,
  and particularly David Dotson and Gareth Shannon who provided in-depth
  comments and suggestions.\par}

\newcommand{\BIBdecl}{\setlength{\itemsep}{0.25 em}}
\bibliographystyle{IEEEtran}
\bibliography{extasy} 
\end{document}